\documentclass[runningheads]{llncs}
\usepackage{graphicx}
\usepackage{tabularx}
\begin{document}

\title{The OpenCitations Data Model}

\author{Marilena Daquino\inst{1,2}\orcidID{0000-0002-1113-7550} \and
Silvio Peroni\inst{1,2}\orcidID{0000-0003-0530-4305} \and
David Shotton\inst{2,3}\orcidID{0000-0001-5506-523X} \and
Giovanni Colavizza\inst{4}\orcidID{0000-0002-9806-084X} \and
Behnam Ghavimi\inst{5}\orcidID{0000-0002-4627-5371} \and
Anne Lauscher\inst{6}\orcidID{0000-0001-8590-9827} \and
Philipp Mayr\inst{5}\orcidID{0000-0002-6656-1658} \and
Matteo Romanello\inst{7}\orcidID{0000-0002-7406-6286} \and
Philipp Zumstein\inst{8}\orcidID{0000-0002-6485-9434}\thanks{
Authors' contributions specified according to the CrediT taxonomy. MD: Conceptualization, Data curation, Formal analysis, Investigation, Methodology, Software, Writing – original draft, Writing – review \& editing. She is responsible for section "Background" and section "Analysis of OCDM reusability". SP and DS: Conceptualization, Investigation, Methodology, Software, Data curation, Supervision, Funding acquisition, Project administration, Writing – original draft. GC, BG, AL, PM, MR and PZ: Investigation, Resources, Validation, Writing – review \& editing.
}}
\authorrunning{M. Daquino et al.}

\institute{Digital Humanities Advanced research Centre (/DH.arc), Department of Classical Philology and Italian Studies, University of Bologna \email{\{marilena.daquino2,silvio.peroni\}@unibo.it} \and
Research Centre for Open Scholarly Metadata, Department of Classical Philology and Italian Studies, University of Bologna \and
Oxford e-Research Centre, University of Oxford
\email{david.shotton@opencitations.net}\\ \and
Institute for Logic, Language and Computation (ILLC), University of Amsterdam\\ \email{g.colavizza@uva.nl}\\ \and
Department of Knowledge Technologies for the Social Sciences, GESIS - Leibniz-Institute for the Social Sciences\\ \email{ghavimi.behnam@gmail.com, philipp.mayr@gesis.org} \and
Data and Web Science Group, University of Mannheim \email{anne@informatik.uni-mannheim.de} \and
École Polytechnique Fédérale de Lausanne\\ \email{matteo.romanello@epfl.ch} \and
Mannheim University Library, University of Mannheim \email{philipp.zumstein@bib.uni-mannheim.de}
}

\maketitle              

\begin{abstract}
A variety of schemas and ontologies are currently used for the machine-readable description of bibliographic entities and citations. This diversity, and the reuse of the same ontology terms with different nuances, generates inconsistencies in data. Adoption of a single data model would facilitate data integration tasks regardless of the data supplier or context application. In this paper we present the OpenCitations Data Model (OCDM), a generic data model for describing bibliographic entities and citations, developed using Semantic Web technologies. We also evaluate the effective reusability of OCDM according to ontology evaluation practices, mention existing users of OCDM, and discuss the use and impact of OCDM in the wider open science community.

\end{abstract}
\section{Introduction}\label{introduction}
In recent years, largely thanks to the Initiative for Open Citations (I4OC)\footnote{\url{https://i4oc.org}}, most major scholarly publishers have made their bibliographic reference data open, resulting, for example, in more than 700 million citations now being made openly available in the OpenCitations Index of Crossref open DOI-to-DOI citations (COCI) \cite{heibi2019software}. As a consequence, scholarly data providers and bibliometric analysis software have started to integrate open citation data into their services, thereby offering an alternative to the current reliance on proprietary citation indexes.

Open bibliographic and citation metadata are beneficial because they enable anyone to perform meta-research studies on the evolution of scholarly knowledge, and allows national and international research assessment exercises characterized by transparent and reproducible processes. Within this context, bibliographic citations are essential components of scholarly discourse, since they “remain the dominant measurable unit of credit in science” \cite{fortunato2018science}. They carry evidence of scholarly networks and of the progress of theories and methods, and are fundamental aids in tenure evaluation and recommendation systems. 
To perform open bibliometric research and analysis, the publications upon which the work is based should be FAIR, namely Findable, Accessible, Interoperable, and Reusable \cite{wilkinson2016fair}. Ideally, such data should be made available without any restrictions, licensed under a Creative Commons CC0 waiver\footnote{\url{https://creativecommons.org/publicdomain/zero/1.0/legalcode}}, and the software for programmatically accessing and analysing them should be also released with open source licences. 

However, data suppliers use a variety of licenses, technologies, and vocabularies for representing the same bibliographic information, or use ontology terms defined in the same ontologies with different nuances, thereby generating diversity in data representation. The adoption of a common, generic, open and documented data model that employs clearly defined ontological terms would ensure data consistency and facilitate integration tasks.

In this paper we present the OpenCitations Data Model (OCDM), a data model based on existing ontologies for describing information in the scholarly bibliographic domain with a particular focus on citations. OCDM has been developed by OpenCitations \cite{peroni2020opencitations}, an infrastructure organization for open scholarship dedicated to the publication of open bibliographic and citation data using Semantic Web technologies. Herein, we propose a holistic approach for evaluating the reusability of OCDM according to ontology evaluation methodologies, and we discuss its uptake, impact, and trustworthiness.  

We compared OCDM to similar existing solutions and found that, to the best of our knowledge, OCDM (a) has the broadest vocabulary coverage, (b) is the best documented data model in this area, and (c) has already a significant uptake in the scholarly community. The main advantages of OCDM, in addition to the consistency of data description that it facilitates, are that it was designed from the outset to enable use by those who are not Semantic Web practitioners, as well as by those that are, that it is properly documented, and it is provided with accompanying software for managing the entire life-cycle of data created according to OCDM.

The paper is organized as follows. In Section \ref{background} we clarify the scope and motivations for this work. In Section \ref{ocdm} we present the data model and its documentation, software and current early adopters. In Section \ref{reusability} we present the criteria we have used to evaluate OCDM reusability and we present results, including figures about OCDM views, downloads and citations according to Figshare and Altmetrics, which are further discussed in Section \ref{discussion}.

\section{Background}\label{background}

The OpenCitations Data Model (OCDM) \cite{peroni2018opencitations} was initially developed in 2016 to describe the data in the OpenCitations Corpus (OCC). In recent years OpenCitations has developed other datasets while OCDM has been adopted by external projects, and OCDM has been expanded to accommodate these changes. We have recently further expanded the OpenCitations Data Model to accommodate the extended metadata requirements of the Open Biomedical Citations in Context Corpus project (CCC). This project has developed an exemplar Linked Open Dataset that includes detailed information on citations, in-text reference pointers such as “Berners-Lee et al. 2011”, and identifiers of the citation contexts (e.g. sentences, paragraphs, sections) within which in-text reference pointers are located, to facilitate textual analysis of citation contexts. 
The citations are treated as first-class data entities \cite{ocidefinition}, enriched with open bibliographic metadata released using a CC0 waiver that can be mined, stored and republished. This includes identifiers specifying the specific positions of the various in-text reference pointers within the text. However, the literal text of these contexts are not stored within the Open Biomedical Citations in Context Corpus, and regrettably in many cases the full text of the published entities cannot be mined from elsewhere in an open way, even for some (view only) Open Access articles, because of copyright, licensing and other Intellectual Property (IP) restrictions.  

Table~\ref{tab1} shows the representational requirements (hereinafter, for the sake of simplicity, also called citation properties and numbered (P1-P8)) that we were interested in recording for each citation instantiated from within a single paper.

\begin{table}[!h]
\centering
\begin{tabularx}{\columnwidth}{|l|X|}
\hline
\textbf{ID} &  \textbf{Description} \\
\hline
P1 & A classification of the type of citation (e.g. self-citation). \\
\hline
P2 & The bibliographic metadata of the citing and cited bibliographic entities (e.g. type of published entity, identifiers, authors, contributors, publication date, publication venues, publication formats). \\
\hline
P3 & The bibliographic reference, typically found within the reference list of the citing bibliographic entity, that references a cited bibliographic entity. \\
\hline
P4 & The separate identifiers of all the in-text reference pointers included in the text of the citing entity, that denote bibliographic references within the reference list. \\
\hline
P5 & The co-occurrence of in-text reference pointers within each in-text reference pointer lists (e.g. “[3,5,12]”). \\
\hline
P6 & The identifiers of structural elements (e.g. XPath of sentences, paragraphs, captions) that specify where, in the full text, an in-text reference pointer appears.\\
\hline
P7 & The function or purpose of the citation (e.g. to cite as background, extend, or agree with the cited entity) to which each in-text reference pointer relates.\\
\hline
P8 & Provenance information of the citation extraction process (e.g. responsible agents, data sources, extraction dates).\\
\hline
\end{tabularx}
\caption{Representational requirements of the OpenCitations Data Model}\label{tab1}
\vspace{-25pt}
\end{table}

\section{The OpenCitations Data Model}\label{ocdm}

\begin{figure}[!h]
    \centering
    \includegraphics[scale=0.245]{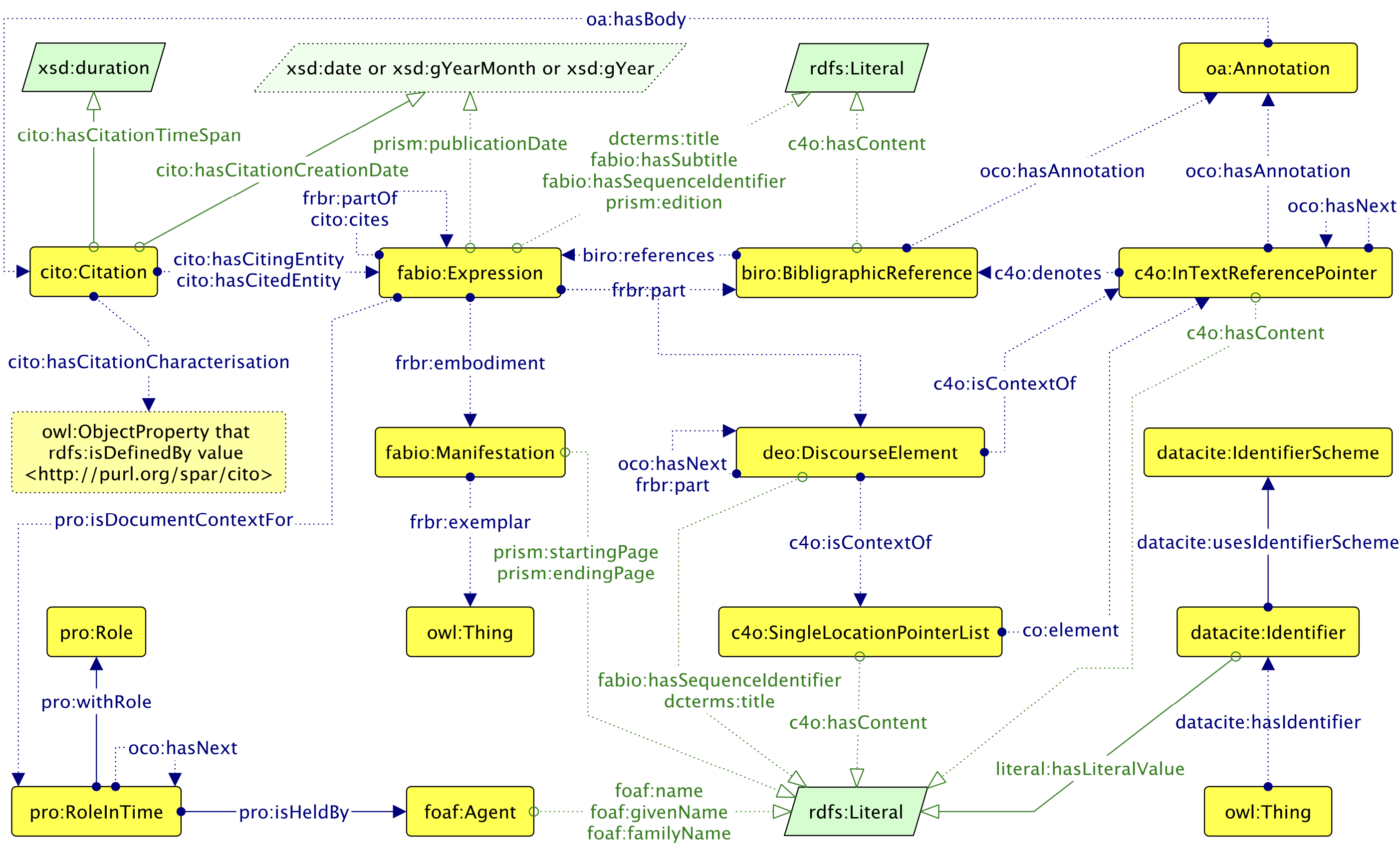}
    \caption{Main classes and properties of the OpenCitations Ontology }
    \label{fig:ocdm}
\end{figure}

The OCDM permits one to record metadata about bibliographic references and their textual contexts, bibliographic entities (citing and cited publications) and the citations that link them, agents and their roles (e.g. author, editor), identifiers for the foregoing entities, provenance metadata and much more, as shown diagrammatically in Fig. \ref{fig:ocdm}.  All terms described in the OCDM are brought together in the OpenCitations Ontology (OCO)\footnote{\url{https://w3id.org/oc/ontology}}. OCO aggregates terms from the SPAR (Semantic Publishing and Referencing) Ontologies \cite{peroni2018spar} and other well-known ontologies, such as PROV-O \cite{belhajjame2013prov} and Web Annotation Ontology \cite{sanderson2013designing}.

Citations are instances of the class \verb|cito:Citation| defined in CiTO, the Citation Typing Ontology\footnote{\url{http://purl.org/spar/cito}}. Subclasses (not shown in Fig. \ref{fig:ocdm}), relevant for P1, include \verb|cito:AuthorSelfCitation|, \verb|cito:JournalSelfCitation|, \verb|cito:FunderSelfCi|\-\verb|tation|, \verb|cito:AffiliationSelfCitation|, and \verb|ci|\-\verb|to:AuthorNetworkSelfCita|\-\verb|tion|. In addition, citations can be characterized with a purpose or function with respect to the related citation context, by means of the property \verb|cito:hasCita|\-\verb|tionCharacterisation| and the use of one or more CiTO properties (e.g. \verb|cito:|\-\verb|usesMethodIn|) (P7).

Instances of the class \verb|fabio:Expression|, defined in the FRBR-aligned Bibliographic Ontology (FaBiO)\footnote{\url{http://purl.org/spar/fabio}}, can be linked to bibliographic metadata such as publication dates, authors, and venues. Instances of \verb|fabio:Manifestation| aggregate information on specific editions and formats (P2).

Instances of \verb|oa:Annotation|, defined in the Web Annotation Ontology (OA)\footnote{\url{https://www.w3.org/ns/oa}}, link instances of the class \verb|cito:Citation| to instances of \verb|biro:Bibliographic|\-\verb|Reference| (P3), defined in BiRO, the Bibliographic Reference Ontology\footnote{\url{http://purl.org/spar/biro}}, and individuals of \verb|c4o:InTextReferencePointer| (P4), defined in C4O, the Citation Counting and Context Characterisation Ontology\footnote{\url{http://purl.org/spar/c4o}}. Lists of in-text reference pointers are represented by the class \verb|c4o:SingleLocationPointer|\-\verb|List| (P5). 

Structural elements wherein in-text reference pointers appear are represented as individuals of \verb|deo:DiscourseElement|, defined in DEO, the Discourse Element Ontology\footnote{\url{http://purl.org/spar/deo}}. Elements are uniquely identified (P6) by means of instances of \verb|datacite:Identifier|, defined in the DataCite Ontology\footnote{\url{http://purl.org/spar/datacite}}.

\begin{figure}[!h]
    \centering
    \includegraphics[scale=0.20]{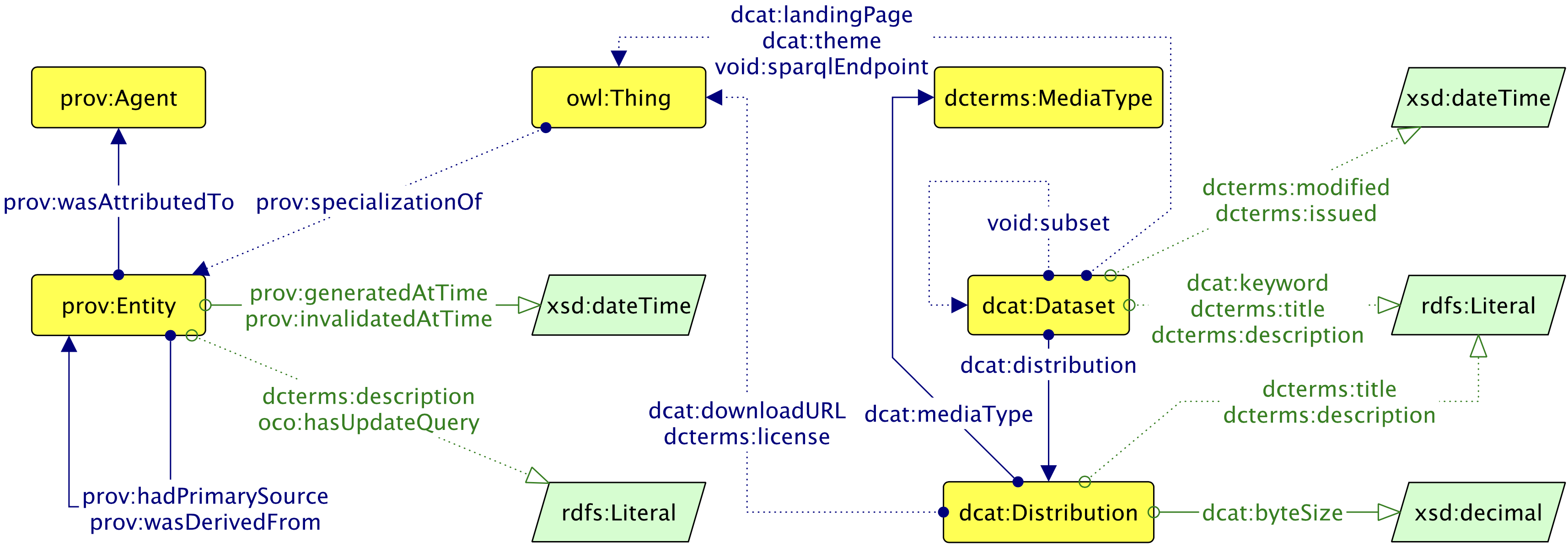}
    \caption{Provenance, versioning, and dataset description in the OCDM}
    \label{fig:prov}
\end{figure}

Finally, as summarized in Figure ~\ref{fig:prov}, OCDM provides guidance for describing the provenance and versioning of each entity under consideration, and also enables the specification of the main metadata related to the datasets containing such entities (P8). To this end, the OCDM reuses terms from PROV-O, the Provenance Ontology\footnote{\url{http://www.w3.org/ns/prov}}, VoID, the Vocabulary of Interlinked Datasets\footnote{\url{http://rdfs.org/ns/void}} \cite{alexander2009describing}, and DCAT, the Data Catalog Vocabulary\footnote{\url{http://www.w3.org/ns/dcat}} \cite{maali2014data}.

Each bibliographic entity described by the OCDM is annotated with one or more provenance snapshots (i.e. instances of \verb|prov:Entity|, each snapshot intended as a specialisation of the bibliographic entity via \verb|prov:specialization|\-\verb|Of|) as defined in \cite{peroni2016document}. In particular, each snapshot records the set of statements having the bibliographic entity as its subject at a fixed point in time, validity dates, responsible agents for either the creation or the modification of the metadata, primary data sources, and a SPARQL query summarising changes with respect to any prior snapshot.

Lastly, a dataset (\verb|dcat:Dataset|) containing information about the bibliographic entities is described with cataloguing information (e.g. title, description, publication and change dates, subjects, webpage, SPARQL endpoint) and distribution information (\verb|dcat:Distribution|) which also includes the specification of licenses, dumps, media types, and data volumes.

\subsection{OCDM documentation and resources}

In order to make the OCDM understandable and reusable by both the Semantic Web community and communities with no expertise in Semantic Web technologies, support material has been produced. All materials are available at \url{http://opencitations.net/model} and include the following resources.

\begin{figure}
    \centering
    \includegraphics[scale=0.45]{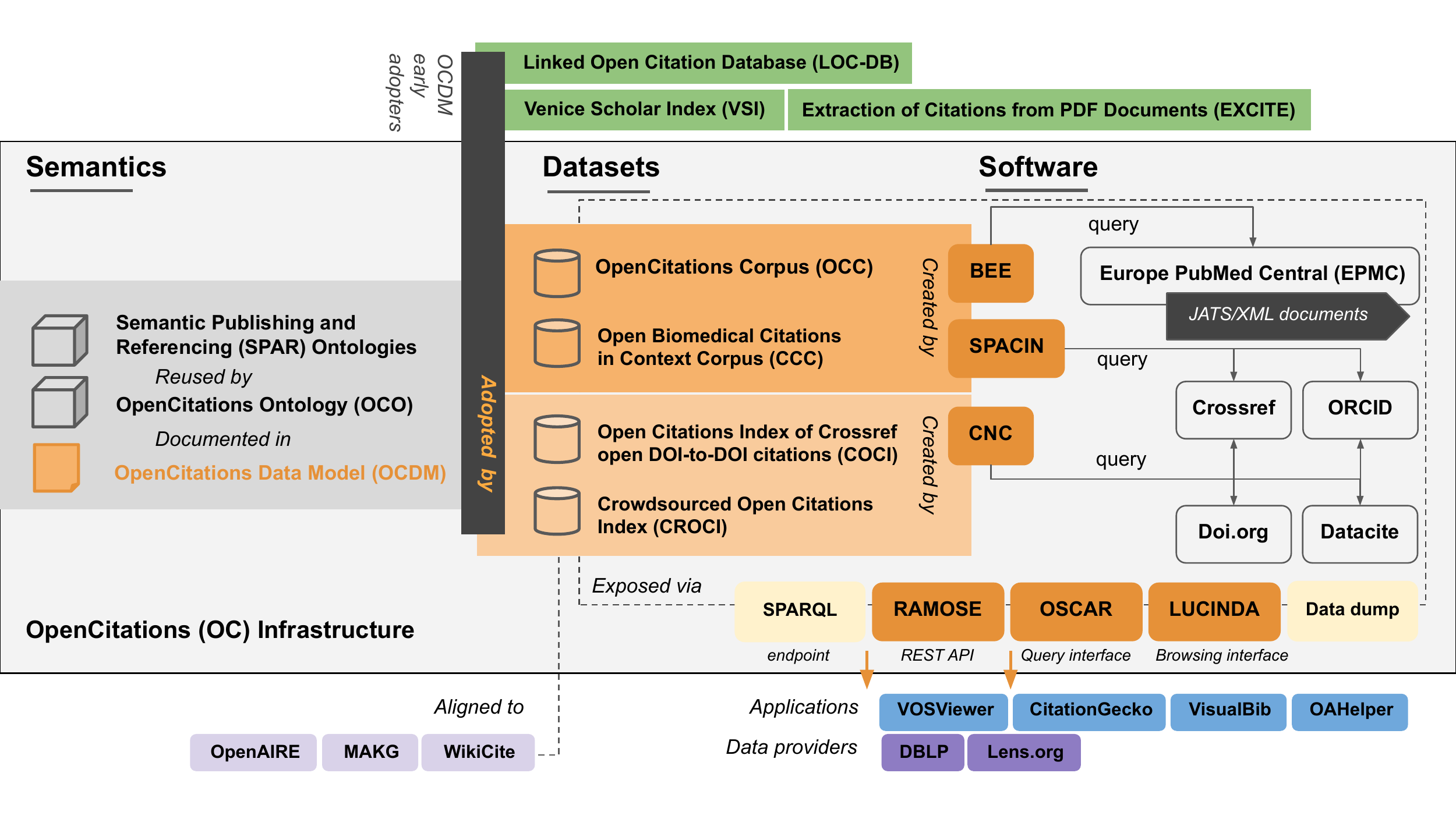}
    \caption{Overview of OpenCitations ecosystem and acronyms used in this paper}
    \label{fig:ocdm_overview}
\end{figure}

\textbf{Human-readable documentation.} The OCDM documentation \cite{peroni2018opencitations} provides (1) detailed definitions of terms characterising open citation data and open bibliographic metadata, (2) naming conventions and URI patterns, and (3) real-world examples. OCDM is supplemented by two additional specifications, i.e. the definition of the Open Citation Identifier (OCI) \cite{ocidefinition} and the definition of the In-Text Reference Pointer Identifier (InTRePID) \cite{intrepid}.

\textbf{OCDM-compliant data examples.} All the data introduced in the OCDM documentation are expressed and provided in JSON-LD to make it easily understandable both to RDF experts and other Web users. In addition, CSV templates have been adopted so as to express and share parts of the OCDM – e.g. to store the citation data in COCI \cite{heibi2019software}.

\textbf{Ontology development documentation.} The first version of the OCDM, released in 2016, addressed citation properties P1-P3 and P8, by directly reusing the SPAR Ontologies and other vocabularies \cite{peroni2018spar}. Within the context of the CCC project described above, we used SAMOD \cite{peroni2016simplified}, an agile data-driven methodology for ontology development, to extend OCO with terms relevant to P4-P7. Motivating scenarios, competency questions, and a glossary of terms of all the new entities included in the OCDM, are available for reproducibility purposes. 

\textbf{Open source software leveraging the data model.} The source code of the knowledge extraction and data re-engineering pipeline for managing data according to OCDM is available at \url{http://opencitations.net/tools}. The pipeline includes software originally developed for creating the OpenCitations Corpus (BEE and SPACIN) and the OpenCitations Indexes (Create New Citations – CNC), and a user-friendly web application (BCite)\cite{demidova2018creating} for creating OCDM-compliant RDF data from lists of bibliographic references. In addition, we have released tools to support the development of applications leveraging data organized according to OCDM: RAMOSE (to create RESTful APIs over SPARQL endpoints), OSCAR (to create user-friendly search interfaces for querying SPARQL endpoints \cite{heibi2017oscar}) and LUCINDA (a configurable browser for RDF data). Configuration files for setting up these tools are available in their GitHub repositories.

\textbf{Licenses for reuse.} OCDM (both the documentation and OCO) is released under a CC-BY license. Software solutions are released under the ISC license. The OCDM-compliant data served by OpenCitations are made open under CC0.

\subsection{OCDM early adopters}

To date, OCDM is central to the work of OpenCitations. The OpenCitations datasets modelled using OCDM include: the OpenCitations Corpus (OCC), including about 13 million citation links and the OpenCitations Indexes, which include more than 721 million citations. Forthcoming datasets, that will be released later in 2020, include OpenCitations Meta, which stores metadata of the citing and cited entities involved in the citations included in the Indexes, and the Open Biomedical Citations in Context Corpus (CCC), mainly derived from the Open Access corpus of biomedical articles provided by PubMed Central, that will include detailed information on in-text reference pointers denoting each reference in the reference list, and their textual contexts. 

Moreover, OCDM has three external acknowledged early adopters. The Extraction of Citations from PDF Documents (EXCITE) project \cite{hosseini2019excite} is run by GESIS and the University of Koblenz. The aim of EXCITE is to extract and match citations from social science publications. To date, EXCITE has extracted around 1 million citations, has converted the data to RDF according to OCDM, and has then published it by ingestion into the OCC.

The Linked Open Citation Database (LOC-DB) \cite{lauscher2018linked} is a project which aims to demonstrate that it is possible for academic libraries to catalogue citation relations sustainably, accurately, and cooperatively. So far, the project has stored bibliographic and citation data for about 7000 published entities. LOC-DB has used a customisation of the OCDM as the data model for defining its data, and exports data in OCDM/JSON-LD so as to be ingested into the OCC.

The Venice Scholar Index (VSI)\footnote{\url{https://scholarindex.eu/}} is an instance of the Scholar Index, originated from the “Linked Books” project \cite{colavizza2019citation} founded by the Swiss National Science Foundation. The citation index includes about 4 million references to publications cited in the historiography of Venice. VSI exports data into RDF formats according to OCDM so as to be integrated into the OCC.

\section{Analysis of OCDM reusability}\label{reusability}

A holistic approach has been used to evaluate the OCO ontology and to infer properties relevant to OCDM. We adopted seminal definitions and classifications of ontology evaluation approaches \cite{brank2005survey,gomez2004ontology} and we selected the following dimensions and approaches that are representative with respect to OCDM reusability. 

\textbf{[E1] Lexical keyword similarity.} This addresses the similarity of definitions (labels of terms) in OCO with respect to the real-world knowledge to be mapped. We adopted a data-driven evaluation \cite{brewster2004data} to map OCO definitions with terms included in a corpus of documents encoded in the Journal Article Tag Suite (JATS) XML schema\footnote{\url{https://jats.nlm.nih.gov/}}. JATS is used by Europe PubMed Central (EPMC)\footnote{\url{https://europepmc.org/downloads/openaccess}} to encode scholarly documents, that are in turn harvested by OpenCitations.

\textbf{[E2] Vocabulary coverage.} This addresses the coverage of concepts, instances, and facts of OCO with respect to the domain to be covered. \textbf{[E2.1]} We validated OCO coverage by comparing it with competing ontologies \cite{maedche2001comparing}. \textbf{[E2.2]} Secondly, we adopted an application-based approach \cite{porzel2004task} to address OCO coverage in four sources that leverage it: OpenCitations, EXCITE, LOC-DB, and ScholarIndex.

Also, we addressed aspects peculiar to OCDM reusability, namely:

\textbf{[E3] Usability-profiling.} This encompasses the communication context of OCDM, i.e. its pragmatics. We evaluated OCDM recognition level \cite{gangemi2006modelling}, i.e. the efficiency of access to OCDM ontologies, documentation, and software, by comparing it with competing ontologies \cite{maedche2001comparing}. 

Lastly, we addressed current uptake, potential impact, and trustworthiness of OCDM, including metrics about OCDM views, downloads and citations according to Figshare and Altmetrics.\footnote{Source code and results of this analysis are available at \url{https://github.com/opencitations/metadata}} 

\subsection{E1: Lexical keyword similarity}

We created a randomized corpus of 2800 JATS documents taken from the Open Access Subset of biomedical literature hosted by Europe PubMed Central. We extracted the list of XML elements used in the documents within this corpus (117 elements), and we expanded element names with definitions scraped from the online XML schema guidelines (e.g. \verb|<p>| became “Paragraph”). We manually pruned non-relevant elements such as MathML markup, text style elements (e.g. \verb|<italic>|), redundant wrapping elements (\verb|<keywordGroup>|) and elements that are out of scope (e.g. \verb|<biography>|), resulting in a refined list of 45 terms.

Secondly, we extracted definitions from OCO (118). We manually pruned terms that were not relevant (e.g. annotation properties, provenance, and distribution related terms), terms that represent hierarchy, sequences, and linguistic aspects not available in XML (e.g. “partOf”, “hasNext”, “Sentence”), and terms dependent on post-processing activities (e.g. “self-citation”, “hasCitationCharacterisation”), resulting in a refined list of 77 OCO definitions.

We then used Wordnet\footnote{\url{https://wordnet.princeton.edu/}} to automatically expand both XML and ontology definitions with synonyms, and we matched synsets similarities. We used a symmetric similarity score to find best matches between the synsets. We considered two thresholds for the similarity match, 0.7 and 0.5, and we manually computed precision and recall. Table~\ref{precision} shows the results.

The coverage of JATS terms in OCO was 55.5\% when the threshold was greater than 0.7, with high precision (96\%) and average recall (53.3\%). The coverage was 73.3\% when the threshold was greater than 0.5, with still high precision (93.3\%) and average recall (68.8\%). 
False negative results included acronyms (e.g. “issn”) that did not have a match in Wordnet, and terms of the taxonomy that were underrepresented in the corpus (e.g. “book”). Likewise, false positive results were due to acronyms used in XML definitions that were not correctly parsed (e.g. “URI for This Same Article Online” was incorrectly matched with “fabio:JournalArticle”).

\begin{table}[]
\centering
\begin{tabularx}{\columnwidth}{|l|X|X|X|}
\hline

\textbf{Threshold} & \textbf{Matches} & \textbf{Precision} & \textbf{Recall}\\
\hline
0.7 & (25/45) 55.5\% & (24/25) 96\% & (24/45) 53.3\%\\
\hline
0.5 & (33/45) 73.3\% & (31/33) 93.9\% & (31/45) 68.8\% \\
\hline
\end{tabularx}
\caption{Lexical similarity between JATS/XML elements and OCO terms}\label{precision}
\vspace{-25pt}
\end{table}

\subsection{E2: Vocabulary coverage}

\textbf{[E2.1] Vocabulary coverage in existing vocabularies.} Since gold standard ontologies are not available, we referred to existing data models and relevant ontologies used by citation data providers. For the sake of completeness, we addressed both open and non-open citation data providers\footnote{See the definition of ``open'' at \url{https://opendefinition.org/licenses/}.}, and both graph data providers and others. We reviewed the vocabulary coverage with respect to P1-P8. We did not take into account discipline coverage or citation counting. The complete list of data models and references is available at \url{https://github.com/opencitations/metadata}. Table~\ref{vocabs} summarizes the comparison of vocabularies coverage, an “x” indicating that the source had metadata of relevance to the citations properties P1-P8 (Table~\ref{tab1}).

\begin{table}[]
\fontsize{8}{9}\selectfont
\begin{tabularx}{\columnwidth}{|l|X|X|X|X|X|X|X|X|}
\hline
 & \textbf{P1} & \textbf{P2} & \textbf{P3} & \textbf{P4} & \textbf{P5} & \textbf{P6} & \textbf{P7} & \textbf{P8}\\
\hline
Google Scholar & & x & & & & & & x\\
\hline
Scopus & & x& & & & & & x\\
\hline
Web of Science & x& x& x& & & & & x\\
\hline
CiteseerX & x& x& x& & & x& &x\\
\hline
Dimensions& & x& x& & & & &x\\
\hline
Crossref& & x& x& & & & &x\\
\hline
EPMC& & x& x& & & & &x\\
\hline
Datacite& &x &x & & & &x &x\\
\hline
DBLP & & x& & & & & &x\\
\hline
MAKG& & x& x& & & x& &\\
\hline
ORC & &x & & & & & &x\\
\hline
GORC& &x &x &x &x &x & &x\\
\hline
SciGraph& &x & & & & & &x\\
\hline
WikiCite& &x & & & & & &x\\
\hline
OpenCitations&x &x &x &x &x &x &x &x\\
\hline
\end{tabularx}
\caption{Vocabulary coverage in existing vocabularies according to P1-P8}\label{vocabs}
\vspace{-25pt}
\end{table}

Non-open citation data providers include Google Scholar, Scopus \cite{scopus}, Web of Science (WoS) \cite{wos}, CiteSeerX \cite{li2006citeseerx} and Dimensions \cite{herzog2020dimensions}. Their data models cover a few aspects of bibliographic metadata (P2) and provenance data (P8). WoS, CiteSeerX, and Dimension also includes bibliographic references (P3). In addition, Wos and CiteSeerX also cover types of citations (P1), and only CiteSeerX includes citation context sentences (P6).

Open citation data providers include Crossref \cite{hendricks2020crossref}, Europe PubMed Central (EPMC), DataCite, DBLP, Microsoft Academic Knowledge Graph (MAKG) \cite{makg} (which is based on Microsoft Academic Graph \cite{wang2020microsoft} and which reuses the SPAR Ontologies and links to resources in Wikidata and OpenCitations), the Semantic Scholar Open Research Corpus (ORC) \cite{ammar2018orc}, the Semantic Scholar’s Graph of References in Context (GORC) \cite{lo2019gorc}, Springer Nature’s SciGraph \cite{scigraph} (which is based on Schema.org),  WikiCite (which includes terms aligned to SPAR Ontologies and interlinks with the OpenCitations Corpus), and the  OpenCitations datasets \cite{peroni2020opencitations}. All data models cover P2, and all except MAKG also cover P8. Only OpenCitations covers P1. In addition, Crossref, Europe PMC, DataCite, MAKG, GORC, and  OpenCitations cover P3. MAKG, GORC, and OpenCitations cover P6, while the latter two also includes in-text reference pointers (P4) and related lists (P5). DataCite and OpenCitations allow the tracking of citation functions (P7).

\textbf{[E2.2] Vocabulary coverage in early adopters.} We separately analysed the vocabulary coverage in acknowledged adopters of OCDM (Table \ref{vocabs_adopters}).

\begin{table}[]
\fontsize{8}{9}\selectfont
\begin{tabularx}{\columnwidth}{|l|X|X|X|X|X|X|X|X|}
\hline
 & \textbf{P1} & \textbf{P2} & \textbf{P3} & \textbf{P4} & \textbf{P5} & \textbf{P6} & \textbf{P7} & \textbf{P8}\\
\hline

\hline
EXCITE& &x &x & & & & &x\\
\hline
LOC-DB& &x &x & & & & &x\\
\hline
VSI& &x &x &x & &x & &x\\
\hline
\end{tabularx}
\caption{Vocabulary coverage in OCDM early adopters according to P1-P8}\label{vocabs_adopters}
\vspace{-15pt}
\end{table}

EXCITE data fully covers P2, P3 and P8. Its local data model also includes information about the data quality of extracted references, which is not currently mapped to OCDM. LOC-DB data fully covers P2, P3, and P8. The OCDM was extended in its local data model so as to cover information about its OCR activities performed on PDF scans. Venice Scholar Index (VSI) aligned data to OCDM terms so as to fully cover P2, P3, P4, P6, and P8. In order to cover peculiar needs of the project relevant to P2, the classes \verb|fabio:Work| and \verb|fabio:Expression| defined in the SPAR Ontologies (and reused in OCO) were specialized so as to include the following sub-classes: \verb|fabio:ArchivalRecord|, \verb|fabio:ArchivalRecordSet|, \verb|fabio:ArchivalDocument|, and \verb|fabio:Archival|\-\verb|DocumentSet|\footnote{As documented at https://github.com/SPAROntologies/fabio/issues/1.}.

\subsection{E3: Usability profiling}
We compared the documentation available for existing graph data providers, namely: MAKG, OC and GORC (Semantic Scholar), SciGraph, and WikiCite. We considered the same dimensions used to address OCDM documentation, namely: human-readable documentation, machine-readable data model and examples, ontology development documentation, open source software leveraging the model, and licenses for reuse (see Table \ref{usability}).

\begin{table}[]
\fontsize{8}{9}\selectfont
\begin{tabularx}{\columnwidth}{|l|X|X|X|X|X|X|X|X|}
\hline
 & \textbf{HR docum.} & \textbf{MR data model} & \textbf{ontology dev. docum.} & \textbf{software} & \textbf{licenses} \\
\hline

\hline
MAKG&x & & & &\\
\hline
ORC and GORC&x &x & & &\\
\hline
SciGraph&x &x & & &x\\
\hline
WikiCite&x & & &x &x\\
\hline
OCDM&x &x &x &x &x\\
\hline
\end{tabularx}
\caption{Usability of existing ontologies and data models}\label{usability}
\vspace{-15pt}
\end{table}

The MAKG data model is graphically represented in \cite{makg}. Software for creating RDF data is available, but no machine-readable data model and examples are provided. Likewise, the development of the data model is not described. Moreover, according to Färber \cite{makg}, the property \verb|c4o:hasContext| is used to annotate instances of \verb|cito:Citation|, rather than \verb|c4o:InTextReferencePointer| as prescribed in C4O, preventing it from representing consistently P3, P4, and P7 in future works, and from merging third-party data with OpenCitations. Lastly, no license is specified for the data model.

The Semantic Scholar Open Research Corpus data model is described in \cite{ammar2018orc}. A machine-readable example of the data model is presented in a dedicated web page\footnote{\url{http://s2-public-api-prod.us-west-2.elasticbeanstalk.com/corpus/}}. No further documentation is available. Similarly, GORC is described in \cite{lo2019gorc}, where an example of JSON data is presented. Both datasets are released under OCD-BY (i.e. an open license), although programmatically accessing data through their APIs requires one to subscribe to a more restrictive and non-open license (comparable to CC-BY-NC-ND). No license associated with the data model is stated.

The Schema.org main classes reused in SciGraph are described in a dedicated web page\footnote{\url{https://scigraph.springernature.com/explorer/datasets/ontology/}}. While the ontology is reused as-is, the SciGraph data model\footnote{\url{https://github.com/springernature/scigraph}} is released as a JSON-LD file and machine-readable examples are available under a CC-BY license. Development documentation of the data model is not available. 

Sources addressing the Wikidata model used by WikiCite include templates\footnote{\url{https://www.wikidata.org/wiki/Template:Bibliographic\_properties}} and examples\footnote{\url{https://www.wikidata.org/wiki/Wikidata:WikiProject\_Source\_MetaData}}. However, no dedicated documentation nor a machine-readable version of the model having citations as a scope is separately available. Data, software, and the general data model are all released under the CC0 license.

Lastly, OCDM \cite{peroni2018opencitations} is described in dedicated human-readable documentation, including machine-readable data model and examples, available under a CC-BY license. The ontology development documentation and the open source software leveraging the model are available on github (ISC licence). All materials are gathered in the official page of the OCDM data model\footnote{\url{http://opencitations.net/model}}.

\subsection{OCDM uptake, potential impact, and trustworthiness }
We can quantify current uptake of the OCDM documentation by using statistics provided by Figshare and Altmetrics, and the number of users’ views of the model description page in the OpenCitations website. As of 18 August 2020, the Figshare document \cite{peroni2018opencitations} has been viewed 10852 times, downloaded 1508 times, and cited 5 times. 100 tweets from 65 users include links to the document. The web page (\url{http://opencitations.net/model}) dedicated to the model has received 13,844 views from 8,202 unique users since 2018.

We can estimate the potential impact of OCDM by considering (a) different types of possible reuse of the model, (b) the number of current reusers of the data model, (c) projects and applications leveraging data created according to OCDM, and (d) the kind of users of data created according to OCDM. 

In detail, OCDM can be reused ‘as is’, via alignment for interchange purposes, and as a JSON data model for non-Semantic Web users. Currently OCDM is used by OpenCitations for all its datasets, and by the three acknowledged early adopters, namely: EXCITE and LOC-DB, which reuse OCDM ‘as is’, and VSI, which aligned terms to OCDM.  EXCITE data have been ingested in the OpenCitations Corpus, while LOC-DB and VSI data are going to be ingested soon. VOSViewer\footnote{\url{https://www.vosviewer.com/}}, CitationGecko\footnote{\url{https://citationgecko.com/}}, VisualBib\footnote{\url{https://visualbib.uniud.it/en/project/}}, and OAHelper\footnote{\url{https://www.otzberg.net/oahelper/}} are applications that leverage OpenCitations data conforming to OCDM retrieved via the OpenCitations REST APIs or directly through its SPARQL endpoints. Moreover, OpenAIRE\footnote{\url{https://www.openaire.eu/}}, MAKG, and WikiCite align data to OpenCitations. Both DBLP and Lens.org\footnote{\url{https://lens.org}} use citation data from OpenCitations to enrich their bibliographic metadata records. 

Users of OpenCitations data include scholars in scientometrics, life sciences, biomedicine, the physical sciences, and the information technology domain. Open\-Citations is currently expanding its coverage to include the social science and the arts and humanities disciplines. The main users of EXCITE data are researchers in the social sciences, while those of the data held by LOC-DB and the Venice Scholar Index include librarians and researchers in the humanities. 

Lastly, we address trustworthiness of OCDM. Long-term availability of ontologies is crucial for the development of the Semantic Web, and the trustworthiness of the ontology creators is important. OCDM, OCO, and the SPAR Ontologies are all maintained by OpenCitations, which has been recently selected by the Global Sustainability Coalition for Open Science Services (SCOSS)\footnote{\url{https://scoss.org/}} as an open infrastructure deserving of crowdfunding support from the scholarly community, thereby helping to ensure its long-term sustainability.

Along with trustworthiness, another important factor is the general interest in the community towards research topics and outputs that can leverage OCDM. So far, two OpenCitations projects dedicated to the enhancement of the OpenCitations Corpus and the creation of the Open Biomedical Citations in Context Corpus have been funded by the Alfred P. Sloan Foundation\footnote{See \url{https://sloan.org/grant-detail/8017}} and the Wellcome Trust respectively, as mentioned above in Section “Background”. Moreover, the Internet Archive and Figshare have both offered to archive backup copies of the OpenCitations datasets without charge.

\section{Discussion and conclusions}\label{discussion}
First, we evaluated lexical similarity of OCO definitions over the knowledge included in data sources encoded in JATS/XML, a gold standard for academic publications [E1]. While the recall is only average, mainly due to mistakes in parsing of acronyms, for those terms that were correctly matched the lexical similarity precision is high, showing that OCO is appropriate for representing data sources organized according to the gold standard. One of the known limits of data-driven evaluation methodologies is that these do not address possible changes in the domain knowledge over time. To date, early adopters of OCO continuously contribute with new scenarios to be represented in the model, which is correspondingly expanded. As a result, OCO will remain a comprehensive reference point for future developments. Other statistical semantic approaches will be evaluated in the future.

Secondly, we evaluated OCO vocabulary coverage as compared with competing data models [E2.1] and in the context of early adopters [E2.2]. Only OCDM fully covers P1-P8. In particular, only one other provider covers P4 and P5 (identifiers for in-text references and groups of these), three providers cover property P6 (although they only store full-text sentences, and lack identifiers for in-text reference pointers), and only one other provider covers property P7 (citation function). Two graph-data providers reuse terms from SPAR Ontologies (either directly or by alignment) in different ways, generating heterogeneity in data. 

Among early adopters, LOC-DB required extensions in order to represent special information related to the cataloguing of digital objects, and VSI required us to expand the FaBiO ontology to permit description of unpublished archival entities. While such changes can be deemed marginal, these are relevant hints for future developments in the humanities domain and will require further analysis. Nonetheless, the OCDM vocabulary coverage is satisfying and strengthens its reusability across domains and applications. 

We showed how alternative citation data providers ensure access to their data models [E3]. Peer-reviewed articles are the main access point to descriptions of those data models, with additional information scattered across various web pages. While machine-readable data models and examples are mostly available, none of the other providers referenced detailed development documentation. Moreover, the licenses for reusing the data models are not always defined. In summary, OCDM appears to be the most documented and findable data model.

Again, no comparison was possible of the uptake of the alternative models in the community. We showed that OCDM has been relatively popular in community social networks, and that the documentation has been downloaded and read by many people. At the moment we cannot measure for what purpose the OCDM documentation has been reused, with the exception of the three early-adopter projects of which we are aware listed in this paper. 

We have shown that OCDM is potentially of significant usefulness to several communities, and fosters reuse in combination with legacy technologies, and we have highlighted ongoing interest from several parties in the maintenance and ongoing development of OCDM in support of several projects.

In future works, we will (a) create SHeX shapes to facilitate reusers in mapping their data to OCDM, and (b) trace OCDM usage scenarios by asking users to fill in a form for statistical purposes. 

\section*{Acknowledgements}
This work was funded by the Wellcome Trust (Wellcome-214471\_Z\_18\_Z). We thank Ludo Waltman (Centre for Science and Technology Studies - CWTS, Leiden University) and Vincent Larivière (École de bibliothéconomie et des sciences de l'information, l’Université de Montréal) for supervising aspects of this work, and Ivan Heibi (University of Bologna) for contributing with suggestions. 

%
%
%
%
\bibliographystyle{splncs04}
\bibliography{bibliography} 

\end{document}